\documentstyle[12pt]{article}

\begin{document}

\font\fourteenbf=cmbx10 scaled\magstep2
%
%
\def\cL{{\cal L}}
%
%
\def\AV{Vachaspati and Ach{\'u}carro}
\def\mn{{\mu\nu}}
\def\a{\alpha}
\def\ga{\gamma}
\def\m{\mu}
\def\n{\nu}
\def\r{\rho}
\def\s{\sigma}
\def\la{\lambda}
\def\k{\kappa}
\def\d{\delta}
\def\dd{\Delta}
\def\t{\tau}
\def\ee{\epsilon}
\def\eps{\varepsilon}
\def\ggg{\Gamma}
\def\om{\omega}
\def\oo{\Omega}
\def\ns{\normalsize}
\def\ssize{\normalsize}
%
%
\def\half{{1 \over 2}}
\def\fourth{{1 \over 4}}
\def\AA{\scriptscriptstyle A}
\def\RR{{\rm I\!\!\, R}}
\def\idx{\int d^4x}
\def\ds{\displaystyle}
%
%

\leftline{CTP\# 2063}
\leftline{DAMTP HEP 92-09}
\leftline{DAMTP R-92/7}
\leftline{NBI-HE 92-08}
\vskip 1.5  true cm

\begin{center}
\Large\bf
Semilocal strings and monopoles
\footnote{This work is supported in part by funds
provided by the U. S. Department of Energy (D.O.E.) under contract
\#DE-AC02-76ER03069.}\\ 
\vspace{0.45in}
\ns\sc G.W. Gibbons\\
\ssize\em
Department of Applied Mathematics and
Theoretical Physics\\
University of Cambridge, Silver Street\\
Cambridge CB3 9EW, UK\\
\vspace{0.2in}
\ns\sc
M.E. Ortiz\\
\ssize\em
Center for Theoretical Physics\\
Laboratory for Nuclear Science
and Department of Physics\\
Massachusetts Institute of Technology\\
Cambridge, Massachusetts 02139, USA\\
\vspace{0.2in}
\ns\sc
F. Ruiz Ruiz\\
\ssize\em
The Niels Bohr Institute\\
University of Copenhagen\\
Blegdamsvej 17\\
DK-2100 Copenahgen \O, Denmark\\
\vspace{0.2in}
\ns\sc
T.M. Samols\\
\ssize\em
King's College\\
Cambridge CB2 1ST, UK\\
and\\
Department of Applied Mathematics and
Theoretical Physics\\
University of Cambridge, Silver Street\\
Cambridge CB3 9EW, UK\\
\vspace{0.8in}
\centerline{Submitted to: {\em Nuclear Physics B}}
\end{center}

\newpage
\vglue 4 true cm

{\leftskip=1.5 true cm \rightskip=1.5 true cm
\openup 1\jot
{\centerline{\bf Abstract}}
\vspace{0.2in}
\noindent
A variation on the abelian Higgs model, with
$SU(2)_{\rm global}\times U(1)_{\rm local}$ symmetry broken to
$U(1)_{\rm global}$, was recently shown by \AV~\cite{tan} to admit
stable, finite energy cosmic string solutions even though the manifold
of minima of the potential energy does not have non-contractible loops.
Here we describe the most general solutions in the Bogomol'nyi limit,
both in the single vortex case and the multi-vortex case.
The single vortex solution depends on one complex parameter and coincides
with that of Hindmarsh \cite{hin}; it may be regarded as a hybrid of a
Nielsen-Olesen vortex and a $CP^1$ lump. The gravitational field of the
vortices considered as cosmic strings is also obtained. Finally,
monopole-like solutions interpolating between a Dirac monopole
and a global monopole surrounded by an event horizon are found.\\
\par}

\newpage
\openup 3\jot
\section{Introduction}

It is often stated that a necessary condition for the existence of
stable cosmic string solutions with finite energy per unit length
is that ``the manifold of minima of the potential energy contains
non-contractible loops''. This condition is not sufficient, as is well
known by now \cite{Raj}. For example, there are no such solutions in a
pure scalar theory with the standard Mexican hat potential.
In a recent paper, \AV~\cite{tan} have constructed a model which
shows that it is not necessary either. Their model has a vacuum manifold
which is topologically $S^3$ but nevertheless admits vortex solutions.
This comes as something of a surprise, since one would
expect that the field at infinity would be free to unwind in $S^3$.
However, it is necessary to take into account the contribution to the
total energy per unit length from spatial gradients.
In order to unwind, the field would have to pass through configurations with
infinite gradient energy.

\AV\ considered the coupling of a two complex component Higgs field
\[
      \Phi=\left({\phi_1\atop\phi_2}\right)
\]
to an abelian gauge field $A_\mu$ via the following matter Lagrangian:
\begin{eqnarray}
      \cL = - \half\> (D_\m \Phi)^{\dag} (D^\m \Phi)
        - {\la \over 8}\,{\left( \Phi^{\dag}\Phi - \eta ^2 \right)}^2
        - \fourth\> F_\mn F^\mn  ~ .
\label{lagr}
\end{eqnarray}
which arises from the standard electroweak
model by setting the $SU(2)$ gauge coupling constant to zero.
Here $D_\m = \partial_\m - i e A_\m$
and the metric has signature $(-,+,+,+)$.
The Lagrangian (\ref{lagr}) has a global
$SU(2)$ symmetry as well as a local $U(1)$ symmetry, under which the
scalar field changes as $\Phi \to e^{i\psi}\Phi$. When the Higgs
field acquires a non-zero vacuum expectation value the symmetry is broken
to a global $U(1)$. The vacuum manifold is the 3-sphere given by
$|\Phi|=\eta$,
which has no non-contractible loops. However, the requirement
that the gradient energy density falls off sufficiently fast further
restricts the field at infinity to lie on a gauge orbit, i.e. a
$U(1)$ orbit of $S^3$, or more precisely a circle lying on $S^3$.

Assuming cylindrical symmetry, a string solution may be found by
making a simplifying ansatz \cite{tan}, equivalent to
setting $\phi_1=f(r)e^{i\theta}$, $\phi_2=0$ and all $A_\mu=0$
except for $A_\theta=A_\theta(r)$. Here $r$ and $\theta$ are the usual
polar co-ordinates on the 2-plane perpendicular to the string. At infinity,
the field configuration is
\[
      \Phi=e^{i\theta}\pmatrix{\eta\cr 0} \>,
\]
provided that $f(r)\to \eta$, which winds once around
a gauge orbit of the vacuum manifold. The equations for $f$ and $A_\theta$
are exactly the equations arising in the Nielsen-Olesen model and so admit
the usual solution, now embedded into the larger semi-local model.
Ref. \cite{tan} deals mainly with the critical limit $\la=e^2$, where the
Higgs mass equals the mass of the gauge field.
In the standard model, this gauge field corresponds
to the field of the $Z$ particle. In the critical limit, the winding
number of the Higgs field around the $U(1)$ orbit at infinity determines the
energy of the solution. Finding solutions of a given energy reduces to solving
the first order Bogomol'nyi equations.

The ansatz above has been generalized by Hindmarsh \cite{hin}.
In the case of critical coupling he finds a family of single string
solutions labelled by a complex parameter. When this parameter vanishes,
one obtains the embedded Nielsen-Olesen solution. When it is non-zero,
the solutions may be regarded as a hybrid of a Nielsen-Olesen vortex and a
$CP^1$ lump. Hindmarsh also studies the stability of these solutions away
from the critical limit.

In this paper, we shall mainly work in the Bogomol'nyi limit $\la=e^2$. In
Sect. 2, we re-examine the flat space Bogmol'nyi equations. Following an
argument employed by Taubes in the standard abelian Higgs model, we derive the
complete set of 2+1 dimensional static multi-vortex (parallel cosmic string)
solutions to these equations, making no further symmetry assumptions.
In the 1-vortex case we recover the solutions described above.
Since these vortices may serve as a model for cosmic strings, we turn
in Sect. 3 to a consideration of their gravitational fields. We give
a Bogomol'nyi bound for the deficit angle in terms of the winding number
and the vacuum expectation value of the Higgs field at infinity.
We also discuss the nature of the solutions of the Einstein-Maxwell-Higgs
equations in the critical limit.

The Lagrangian (\ref{lagr}) does not admit non-singular monopole
solutions in the usual sense. However, it does admit singular monopole
solutions which are a modification of the Dirac monopole by the
additional Higgs fields. The energy density of the Higgs fields falls
off as slowly as it does in the case of a global monopole. This means
that in flat space it would have infinite total energy. When coupled
to gravity it gives rise to an asymptotically conical spacetime. Such
objects are probably of limited interest for particle physics but may
have applications to cosmology. In Sect. 4 we consider the gravitational
field of these monopoles, including the case when the singularity at the
origin is hidden inside a black hole.

\section{Semi-local strings in flat space: the Bogomol'nyi limit}

We begin by examining the static flat space string solutions in the
Bogomol'nyi limit, generalizing the work of \cite{tan} and \cite{hin}.
Imposing translational symmetry, the strings are all parallel, and the
problem reduces to one of vortices in 2+1 dimensions. The corresponding
critically-coupled problem in the standard abelian Higgs model, with
only one Higgs field, has been completely analysed by Taubes \cite{J&T}.
Taubes showed that in the  $n$th  topological sector the solutions are
labelled by the choice of $n$ unordered points in the plane; these are the
points  where the Higgs field vanishes, and for large separations they
may be identified as the positions of the vortices. Here we shall find a
similar space of solutions, though with additional parameters describing
extra degrees of freedom besides position. We shall extensively invoke
Taubes' results in our analysis.

It is convenient to work with dimensionless quantities.  Accordingly,
in this section we make the replacements
\[
  \phi_a \to \eta\phi_a\,,\qquad A_\m \to \eta A_\m\,,\qquad x\to x/e\eta
\]
and introduce the parameter $\a =\la/e^2$. Working in the gauge $A_0=0$,
we seek the stationary points of the static energy functional
\[
     E= \int d^2\!x\, \varrho ~ ,
\]
where the energy density $\varrho$ is given by
\[
     \varrho = - T^0_{~0} = e^2\eta^4\left[\,
       {1\over 2}\,(\overline {D^i \phi_a}) (D_i \phi_a)
       + {1\over 4}\,F_{ij}F^{ij}
       + {\a \over 8}\,(\bar\phi_a\phi_a -1)^2\, \right]
\]
and the index $i=1,2$ labels the coordinates transverse to the string.
$D_i$ is now defined as $\partial_i-iA_i$.
Finiteness of $E$ implies the boundary conditions
\begin{eqnarray}
     \phi_a \bar\phi_a \to 1 \qquad |x| \to \infty
     \label{b1}
\end{eqnarray}
and
\begin{eqnarray}
     D_i \phi_a \to 0 \qquad |x| \to \infty \> ,
     \label{b2}
\end{eqnarray}
where the limits are approached faster than $O(|x|^{-1})$.
The first of these conditions means that on the circle at infinity,
$S^1_{\infty}$, the Higgs field must lie on the 3-sphere,
$S_{\phi}^3=\{\phi:\phi_a\bar\phi_a=1\}$; the second requires
that it varies by at most a pure phase there. Phasing the Higgs
field fibres $S_{\phi}^3$ as a $U(1)$ bundle over $CP^1 \cong S^2$
(the Hopf fibration of $S^3$). A given sector of the theory is
thus specified by a choice of a point on $S^2$ and the winding number of
the Higgs field around the fibre over that point.

The global $SU(2)$ symmetry allows us to assume, without loss of
generality, the particular asymptotic form
\begin{eqnarray}
     \left({\phi_1 \atop \phi_2}\right)
       \to \left({\phi_0 \atop 0}\right)\,,
       \qquad          |\phi_0|=1
     \label{asy}
\end{eqnarray}
at $|x| \to \infty$, which corresponds to a specific choice of fibre.
We denote by $n$ the winding around the fibre, i.e. the winding number
of the map
\[
     \phi_0: S^1_\infty \to U(1) \, .
\]
The total magnetic flux through the plane is then [using (\ref{b2}) and
Stokes's theorem]:
\[
     \oint_{S^1_\infty} A_i dx^i = 2\pi n \, .
\]

The static energy density may be rewritten in the usual way as a sum of
squares, a term proportional to the potential and a total divergence
\cite{bog}:
\begin{eqnarray}
  - T^0_{~0} & = e^2\eta^4 \,\biggl\{
  {\displaystyle 1\over 8}\, {\left[ \eps^{ij} F_{ij}
  \pm (\bar{\phi}_a \phi_a  -1 )\right]}^2
  + {\displaystyle 1\over 4} \, {|D_i\phi_a \pm i\eps_i^{\,j}D_j\phi_a |}^2
  \nonumber
  \\
  & +{\displaystyle 1\over 8}\,(\a -1)\,
  {\left( \bar{\phi}_a \phi_a -1 \right) }^2
 \pm\partial_i(\eps^{ij}J_j)\biggr\} \> ,
 \label{trick}
\end{eqnarray}
with $\eps_{ij}= -\eps_{ji}$, $\eps_{12}=1$ and
\[
     4J_j = 2 A_j - i\bar{\phi}_a D_j\phi_a
       + i \phi_a \overline{D_j\phi_a} ~  .
\]
Specializing to $\a = 1$, one obtains
\[
     E \geq \pi\eta^2 |n|\, ,
\]
with equality if and only if the Bogomol'nyi equations
\begin{eqnarray}
     (D_1 \pm iD_2)\phi_a = 0
     \label{bog1}
\end{eqnarray}
and
\begin{eqnarray}
     F_{12} \pm {1\over 2}(\phi_a \bar\phi_a-1) = 0
     \label{bog2}
\end{eqnarray}
are satisfied, the upper and lower signs corresponding to $n>0$ and
$n<0$ respectively.
Their solutions minimise the static energy, so automatically satisfy
the full second-order static equations. In the standard abelian Higgs
model with only a single Higgs field, Taubes has shown that the
converse is also true \cite{J&T}. A similar proof works here too, so
that to study the static theory it suffices to consider just the
first order Bogomol'nyi equations. In the following we assume $n>0$
and take the upper signs.

Consider first points of the plane at which $\phi_a \ne 0$.
The first equation (\ref{bog1}) may then be rewritten as
\begin{eqnarray}
     A = i \partial_z \ln \bar \phi_a \qquad\qquad a=1,2,
     \label{ac}
\end{eqnarray}
where we have introduced the complex notation $z=x_1+ix_2$ and
$A={1 \over 2}(A_1-iA_2)$.
The difference of these equations yields
\[
     \partial_{\bar z} \ln \left({\phi_2 \over \phi_1} \right)=0
\]
so that the ratio
\begin{eqnarray}
     w(z)={\phi_2 \over \phi_1}
     \label{quo}
\end{eqnarray}
is locally analytic in $z$.
{}From the remaining Bogomol'nyi equation (\ref{bog2}) -- using (\ref{ac})
to eliminate $A_i$, and writing $f=\ln |\phi_1|^2$ -- we obtain:
\begin{eqnarray}
     \triangle f +1 - (1+|w|^2){\rm e}^f = 0 \, .
     \label{feq}
\end{eqnarray}

Now let us consider the zero set of $\phi_1$. As described in \cite{J&T},
an application of the $\bar \partial$-Poincar\'e lemma shows that for
smooth solutions of (\ref{ac}), $\phi_1$ has zeros at discrete points
in the plane, with local behaviour in the neighbourhood of a zero $z_0$
of multiplicity $n_0$ given by
\begin{eqnarray}
     \phi_1(x) = (z-z_0)^{n_0} h(x) \, ,
     \label{z}
\end{eqnarray}
$h$ being a smooth, non-vanishing function of $x$. Moreover, the
boundary condition on $\phi_1$ means that there are
precisely $n$ zeros (counted with multiplicity) \cite{J&T}.
Let us denote these zeros by $z_r$ $(r=1,\ldots n)$.
Since by (\ref{asy}), $w(z)$ vanishes at infinity, it follows that
\begin{eqnarray}
     w(z)= {Q_n(z)\over P_n(z)} \, ,
     \label{qp}
\end{eqnarray}
where
\begin{eqnarray*}
     P_n(z) && =\prod_{r=1}^n (z-z_r)
     \\
     &&\equiv  z^n + p_{n-1} z^{n-1} + \ldots + p_1 z + p_0
\end{eqnarray*}
and
\[
     Q_n(z) \equiv q_{n-1} z^{n-1} + \ldots + q_1 z + q_0
\]
is a polynomial of at most order $n-1$, possibly sharing
roots with $P_n(z)$;
if it does, then $w(z)$ is defined at these points by continuity.
Recalling (\ref{z}), and noting that in two dimensions
\begin{eqnarray}
     \triangle \ln|z-z_r|^2 = 4 \pi \delta (x-x_r) \, ,
     \label{2d}
\end{eqnarray}
we see that the extension of (\ref{feq}) to all points of $\RR^2$ is
\begin{eqnarray}
     && \triangle f+1 - (1+|w|^2){\rm e}^f = 4\pi \sum_{i=1}^n
       \delta (x-x_r) \nonumber
     \\
     && f \to 0  \qquad{\rm as} \qquad |x|\to \infty \, ,
     \label{prob}
\end{eqnarray}
with $w(z)$ as in (\ref{qp}).

Our problem is thus reduced to analysing the solutions of (\ref{prob}).
It is convenient to introduce the quantity
\begin{eqnarray}
     u=f + \ln (1+|w|^2)
     \label{uu}
\end{eqnarray}
so that (\ref{prob}) becomes
\begin{eqnarray}
     \begin{array}{l}
          \triangle u +1 - {\rm e}^u = \rho
          \\
          \\
          u \to 0 \qquad {\rm as} \qquad |x| \to \infty \, ,
     \end{array}
     \label{ueq}
\end{eqnarray}
with ``source'' term
\begin{eqnarray}
     \rho = \triangle \ln (|P_n|^2 + |Q_n|^2) \, .
\end{eqnarray}
The standard abelian Higgs model has $Q_n=0$, hence
$\rho=4\pi \sum_{i=1}^n \delta (x-x_r)$. In this case Taubes has proved the
existence and uniqueness of the solution of (\ref{ueq}) for any choice of the
$x_r$ \cite{J&T}. A minor modification of his work does the same in our more
general problem. In analogy with \cite{J&T}, we define
\[
  u_1 = \ln (|P_n|^2 + |Q_n|^2) - \sum_{r=1}^n \ln (|z-z_r|^2 + \mu) \, ,
\]
where $\mu>0$  is for the moment arbitrary. Setting $u=u_1+v$ then
gives
\begin{eqnarray}
     \begin{array}{l}
          \triangle v + g_0 -  1- {\rm e}^{u_1} {\rm e}^v = 0
          \\
          \\
          v \to 0 \qquad {\rm as} \qquad |x| \to \infty \, ,
     \end{array}
     \label{veq}
\end{eqnarray}
where
\[
     g_0= 4\sum_{r=1}^n {\mu \over (|z-z_r|^2+\mu)^2} \, .
\]
(\ref{veq}) is the variational problem associated with the functional
\[
     a(v)= \int \, d^2x \left[\, |\nabla v|^2 + v(1-g_0) - {\rm e}^{u_1}
       ({\rm e}^v - 1)\, \right] \, .
\]
Taubes' method for $Q_n=0$ rests essentially on showing: (i) that $a(v)$ is
strictly convex, and (ii) that for a large enough ball in function space,
the normal derivative of $a(v)$ on the boundary of the ball is positive;
together (i) and (ii) imply the existence of a unique minimum of $a(v)$.
When $Q_n \ne 0$ these results still hold. For (ii) one must check
the inequalities \cite{J&T}
\begin{eqnarray}
     \left.
     \begin{array}{l}
          1-g_0 > c \qquad {\rm for~some}~c\in (0,1)
          \\
          \\
          1-g_0 - {\rm e}^{u_1} \ge 0
     \end{array}
     \right\}
       {\rm ~~for~all~} x\in \RR^2 \, .
     \label{ineq}
\end{eqnarray}
Noting that
\[
     {\rm e}^{u_1} = {\prod_{r=1}^n |z-z_r|^2 +
       \left|\sum_{k=0}^{n-1} q_k z^k \right|^2
       \over \prod_{r=1}^n  (|z-z_r|^2 + \mu)} \> ,
\]
it is clear that for a given $P_n$ and $Q_n$, the inequalities (\ref{ineq})
can be satisfied for a sufficiently large $\mu$.

We conclude then that for every choice of complex polynomials $P_n$ and $Q_n$,
a unique solution exists. Using (\ref{quo}), (\ref{z}) and (\ref{uu}),
the Higgs field may be reconstructed from $u$ via
\begin{eqnarray}
     \left( \phi_1 \atop \phi_2 \right)
       &&={1\over \sqrt{1+|w|^2}} \left(1 \atop w \right)
       {\rm e}^{{1\over 2} u}
       \prod_{r=1}^n {(z-z_r) \over |z-z_r|}
     \nonumber
      \\
     &&={1\over \sqrt{|P_n|^2+|Q_n|^2}} \left(P_n \atop Q_n \right)
       {\rm e}^{{1\over 2} u}
     \label{result}
\end{eqnarray}
up to gauge transformations\footnote{Note that if $P_n$ and $Q_n$ have
a common root then ${\rm e}^{{1\over 2}u}$ has a zero there, so these
expressions are everywhere well-defined.}. $A_i$ is then given
by (\ref{ac}). The moduli space of solutions (i.e. identifying
gauge equivalent configurations) is thus just $C^{2n}$,
the $4n$-dimensional space parametrised by the coefficients
of $P_n$ and $Q_n$, $\{p_k, q_k:k=0,1,\ldots,n-1\}$.

When $n=1$ we recover the single soliton solutions of
\cite{hin}, and {\it a fortiori} that of \cite{tan}.
The solutions are labelled by the two complex
parameters $p_0=-z_1$ and $q_0$. Let us write
$z-z_1=r{\rm e}^{i\theta}$ so that $(r,\theta)$ are polar co-ordinates
in the plane centred at $z_1$. $u$ is then a function of $r$ only and
(\ref{ueq}) reduces to an ordinary differential equation. The expression
(\ref{result}) for the Higgs field becomes
\[
     \left( \phi_1 \atop \phi_2 \right)
      ={1\over \sqrt{r^2+|q_0|^2}} \left(r{\rm e}^{i\theta} \atop q_0 \right)
      \exp \left\{ {1\over 2}\, u(r;|q_0|) \right\} \> .
\]
This is the form of the ansatz made by Hindmarsh \cite{hin}. It
corresponds to a vortex-like structure centred at $z_1$, with
size and orientation determined by the complex parameter $q_0$.
If $q_0 \ne 0$, the Higgs field is non-zero at $z_1$ and approaches
its asymptotic values like $O(r^{-2})$. In the limit $|q_0| \to 0$,
one recovers the solution in \cite{tan}, i.e. the usual
critically-coupled abelian Higgs vortex, with the Higgs
vanishing at $r=0$ and approaching the vacuum exponentially fast.
On the other hand, when  $|q_0| \gg 1$, the solution approximates
a $CP^1$ lump \cite{hin}. To see this, note that for $|q_0| \gg 1$,
$\rho=4|q_0|^2/(r^2+|q_0|^2)^2 \simeq 0$. The solution of (\ref{ueq})
is therefore $u \simeq 0$ so the Higgs field lies on the
vacuum manifold $S^3$:
\begin{eqnarray}
      \left( \phi_1 \atop \phi_2 \right)
      \simeq{1\over \sqrt{|z-z_1|^2+|q_0|^2}}
       \left(z-z_1 \atop q_0 \right) \, .
\label{higvac}
\end{eqnarray}
As remarked before, the action of the $U(1)$ symmetry fibres this $S^3$
as a circle bundle over $CP^1$ --
the Hopf bundle. Quotienting out by $U(1)$ gauge transformations,
the Higgs field defines a map from the plane $\RR^2 \cong C$
into $CP^1$. By (\ref{higvac}), this map is analytic and of degree 1,
i.e. a $CP^1$ lump. The solitons
of this model thus interpolate between abelian Higgs vortices and
$CP^1$ lumps.

A general $n$-soliton configuration may, at least for separations much
larger than the soliton sizes, be regarded as an approximate superposition
of $n$ 1-solitons. The $4n$ dimensions of the moduli space are accounted for,
roughly speaking, by the positions of the solitons and their sizes and
internal phases. If we fix the scale parameters $q_k$ to be zero, we recover
the usual $2n$-dimensional moduli space of abelian Higgs vortices, $C^n$.

Given the moduli space of static solutions, one is in a position to
understand the scattering of the solitons.
At low energies the dynamics should be well-approximated by
geodesic motion on the moduli space
equipped with the metric induced by the field kinetic energy \cite{man}.
One would expect to find similarities with the corresponding analyses
of abelian Higgs vortices \cite{rub} and lumps \cite{lee}.

The generalization of the results to a
theory with an $p$-component Higgs field, remarked upon in \cite{hin}, is
straightforward. The vacuum manifold is now $S^{2p-1}$. This is a
$U(1)$ bundle over $CP^{p-1}$ and again the asymptotic Higgs
field must wind round one of the fibres.
The expression for the Higgs field is the obvious generalization of
(\ref{result}), and the $n$-soliton moduli space is $C^{pn}$.

\section{Semilocal strings in curved space}

We now turn to a consideration of the gravitational properties
of the solitons of the previous section, regarded as cosmic strings.
We work with the usual Einstein-Hilbert action
\[
     S = {1 \over 16\pi G} \idx ~ \sqrt{-g} ~ R
       + \idx~ \sqrt{-g}~\cL ~ ,
\]
where $\cL$ is the matter Lagrangian (\ref{lagr}) with now all
derivatives taken to be covariant and the metric appearing as a
dynamical variable.

Letting the string lie along the $x^3$-axis and assuming boost
invariance in the $(t,x^3)$-plane, the most general form of the metric
is \cite{8920}
\[
     ds^2 = W^2 \left[ - dt^2 + (dx^3)^2 \right]
       + h_{ij} dx^i dx^j \, ,
\]
where $W$ and $h_{ij}$ are functions of $x^i$ only. Being 2-dimensional,
the metric on the sections transverse to the string admits a K\"ahler
form which we denote $\eps_{ij}$, $\eps_i^{\,j}$
the complex structure. In complex local co-ordinates we have
\[
     \eps_{ij} dx^i \wedge dx^j = i \oo ^2 dz\wedge d\bar{z} \, ,
\]
with $\oo$ the conformal factor,
\[
     h_{ij} dx^i dx^j = \oo^2  dz d\bar{z} \, .
\]
The only non-vanishing components of the Einstein tensor
$G_{\mn}=R_{\mn} - \half g_{\mn}R$ are
\begin{eqnarray}
   && G_{00} = - G_{33} = W^2K - W \nabla^2 W \, ,
     \nonumber
      \\
   && G_{ij} = -2W^{-1}\nabla_i \nabla_j W
       + h_{ij}W^{-2}{(\nabla W)}^2
       + 2h_{ij}W^{-1}\nabla^2 W \, ,
     \label{ein}
\end{eqnarray}
where $K$ is the Gauss curvature of $h_{ij}$ and $\nabla_i$ is the
covariant derivative with respect to $h_{ij}$.

Following the arguments in \cite{C&G}, it can already be shown that
the total deficit angle is positive, provided the energy density is
non-negative. Consider Einstein's equations, which in our conventions
read $G_{\mn} = 8\pi G T_{\mn}$ and look at the $00$-component.
With the help of (\ref{ein}) it can be cast as
\[
     K - W^{-1}\nabla^2 W = - 8\pi G T^0_{~0} \, .
\]
The topology of the transverse sections will in general be that
of a 2-disc so that the Gauss-Bonnet theorem gives
\[
     \int K~ \sqrt{h}~d^2\! x = \d,
\]
with $\d$ the deficit angle. Integrating by parts and discarding a
boundary term, we then conclude that
\begin{eqnarray}
    \d =   \int  \sqrt{h}\, d^2 x \,
    \left(- 8\pi G T^0_{~0}
      + \left| {\nabla W \over W} \right|^2
    \right) \ge  \int  \sqrt{h}\, d^2 x \,
    \left(- 8\pi G T^0_{~0} \right)~ .
    \label{zzz}
\end{eqnarray}
Using that the K\"ahler form $\eps_{ij}$ is covariantly constant,
the energy density can be rearranged as in (\ref{trick}). The only
difference is that now $\eps_{ij}$ involves the metric coefficients
$h_{ij}$ and that instead of the ordinary derivative $\partial_i$ we have
the covariant derivative $\nabla_i.$ It thus follows that for field
configurations with boundary conditions (\ref{b1}-\ref{b2})
and $\a\ge 1,$
\begin{eqnarray}
     \d \ge 8\pi^2 G\eta^2 |n| \, ,
     \label{bo}
\end{eqnarray}
$n$ being the topological number entering in the flux.

{}From (\ref{trick}) and (\ref{zzz}) it follows that for
$\a \ge 1$, the configuration realizing the bound in (\ref{bo}) has
$\a =1$ and satisfies the Bogomol'nyi equations
\begin{eqnarray}
     (D_i \pm i\eps_i^{\,j}D_j)\phi_a = 0\, ,
     \label{1}
\end{eqnarray}
\begin{eqnarray}
     \eps^{ij}F_{ij} \pm(\phi_a \bar\phi_a-\eta^2) = 0
     \label{2}
\end{eqnarray}
and the Einstein equations. From equality in (\ref{zzz}) we deduce
that $W$ is a constant, which can always be taken equal to one by suitably
choosing coordinates $t$ and $x^3$. Using (\ref{1}) and (\ref{2}), the
energy-momentum tensor becomes
\[
     T^0_{~0}=T^3_{~3}= \mp\nabla_i(\eps^{ij} J_j)\, , \quad  T_{ij}=0
\]
so that the Einstein equations take the form:
\begin{eqnarray}
     && K - W^{-1}\nabla^2 W=\pm 8\pi G
       \nabla_i(\eps^{ij} J_j) \, ,
     \label{3}
     \\
     && -2W^{-1}\nabla_i \nabla_j W
       + h_{ij}W^{-2}{(\nabla W)}^2
       + 2h_{ij}W^{-1}\nabla^2 W  = 0 \, .
     \label{4}
\end{eqnarray}

Eq. (\ref{1}) is the same as for flat space; in complex local coordinates
it takes the form (\ref{ac}). From the analysis of the previous section
we know that $w(z)=\phi_2/\phi_1$ is locally holomorphic in $z$
and vanishes at infinity, with $\phi_1$ having $n$ zeros counted with
multiplicity. Thus, eq. (\ref{2}) can be rewritten as
\[
     {1\over {\oo ^2}}\triangle \ln
       {{|\phi_1|^2}\over{\prod_{r=1}^n |z-z_r|^2}}=
        \left(1+|w|^2\right)|\phi_1|^2-\eta^2,
\]
where we have used (\ref{2d}) for the zeros of $\phi_1.$ As for
eq. (\ref{3}), we first note that for points at which $\phi_1\neq 0,$
the density of energy can be cast with the help of (\ref{ac}) as
\[
     -T^0_{~0} ={1\over{4\oo^2}}\triangle
       \left[\left(1+|w|^2\right)|\phi_1|^2
       -\ln |\phi_1|^2 \right] \, .
\]
Secondly, we recall that in conformal coordinates the Gauss
curvature $K$ reads
\[
     K = -{1\over {2 \oo ^2}}\triangle\ln \oo^2 \, .
\]
Eq. (\ref{3}) then becomes
\begin{eqnarray}
     \triangle  \left\{\ln \oo^2+4 \pi G
       \left[\left(1+|w|^2\right)|\phi_1|^2
        -\ln|\phi_1|^2\right]\right\} = 0 \, .
     \label{3p}
\end{eqnarray}
Finally, to extend (\ref{3p}) to the zeros of $\phi_1$ we again use (\ref{2d})
and obtain that
\[
     \ln\oo^2 +4\pi G\left[
       \left(1+|w|^2\right)|\phi_1|^2
       -\ln{{|\phi_1|^2}\over{\prod_{i=1}^n
       |z-z_r|^2}}\right]
\]
is harmonic and bounded, hence a constant. Using now that at infinity
$w \to 0$ and $|\phi_1| \to \eta$, we get the following asymptotic behaviour
for the conformal factor:
\begin{eqnarray}
     \oo\to|z\bar{z}|^{-4\pi Gn\eta^2}\quad{\rm as}
       \quad|z|\to\infty \, .
     \label{conf}
\end{eqnarray}
To see that a 2-metric with conformal factor (\ref{conf}) corresponds to
a conical metric with deficit angle $\d = 8\pi G n \eta^2$ it is enough to
perform the change of variables $|z|^{\a} = \a r$,
${\rm arg}(z) = \theta,$ with $\a = 1 - 4\pi G n \eta^2$ and $r$ and
$\theta$ polar coordinates on the plane.

\section{Semi-local monopoles}

In this section we shall
present  a spherically symmetric  monopole-like solution of our model
coupled to gravity. It is a kind of hybrid structure incorporating
some of the features of global monopoles with some of the features of
Dirac  monopoles. Note that because the theory is abelian,
${\bf\nabla}\!\cdot\! {\bf B}=0$ everywhere if the fields are regular.
Therefore any magnetic monopole must have a singularity somewhere. In
flat spacetime we could simply regard our solution as a large distance
approximation of some more basic theory possessing non-singular
monopoles. When gravity is included, it possesses singularities at the
origin but these singularities may be hidden inside an event horizon
and we shall therefore not investigate them in detail, merely
confining ourselves to checking that our solution is regular outside a
regular event horizon.

We shall assume that the spacetime metric is spherically symmetric and
static outside a regular event horizon. We may therefore take
\[
     ds^2=-C^2(r)dt^2 +D^2(r)dr^2+r^2(d \theta ^2
       +\sin^2\theta d\phi^2) \, .
\]
We now work with the original dimensionful quantities $=\Phi$, $A$ and $x$.
The vector potential $A$ is assumed to take the form:
\[
     A= {1\over{2e}}\cos\theta \,d\phi \, .
\]
This represents the electromagnetic field of a Dirac monopole with the
lowest possible magnetic charge, i.e. the  magnetic charge $g = 2 \pi
/e$. Restricted to the 2-spheres $t = constant$ and  $r=constant$, it
provides a connection on the basic Hopf bundle of Chern class 1 over
the 2-sphere. The total space of this bundle is just the 3-sphere, and
the projection map is just the projection onto the 2-sphere's worth of
Hopf fibres.

For the Higgs field we postulate that
\begin{eqnarray}
     \left( \phi_1 \atop \phi_2 \right)
       = f(r) \left({\rm e}^{i\phi/2} \cos(\theta/2)
        \atop {\rm e}^{-i\phi/2} \sin(\theta/2) \right)~.
     \label{ansatz}
\end{eqnarray}
One readily checks that our ansatz is consistent with the coupled
Einstein-Maxwell-Higgs equations and leads to a set of coupled
non-linear radial equations for the functions $f(r)$, $C(r)$ and $D(r)$.
Note that while the Higgs component $\phi_2$ is well-defined at the
north pole $\theta =0$, it has a string-like singularity at the south
pole $\theta= \pi$. On the other hand, the component $\phi_1$ has a
string-like singularity at the north pole and is well-defined
at the south pole. As for the vector potential, it has Dirac
string singularities at the north and south poles.
However, these two string-like singularities, both in
the Higgs field and the vector potential, are gauge artifacts and may
be shifted by performing a gauge transformation.

In the ``$\sigma$-model limit'' in which the Higgs
field stays in the vacuum manifold everywhere, i.e.
\[
     f(r)  = \eta \, ,
\]
the radial equations for our ansatz (\ref{ansatz}) become trivial and
an exact solution may be obtained. This approximation would hold rigorously
if we took the limit $\lambda \to \infty$. It also holds asymptotically
in the large $r$ limit. This being granted, the metric turns out to be
given by
\begin{eqnarray}
      \begin{array}{l}
      ds^2=-Vdt^2 + V^{-1}dr^2+r^2(d\theta^2+\sin^2
      \theta d\phi^2) \, ,
      \\
      \\
      V=\left(1-2\pi G\eta^2-\ds{{2GM}\over
       r}+\ds{{Q^2}\over{r^2}}\right)
      \quad{\rm and}\quad Q^2 = \ds{{\pi G} \over { e^2}}.
      \end{array}
      \label{metricform}
\end{eqnarray}

Two special cases of this metric are
interesting. One is when no Higgs field is present, i.e.
the case
\[
     f(r) = \eta = 0 \, .
\]
The solution then reduces to the  well known asymptotically flat
Reissner-Nordstr\"om one representing a Dirac monopole with source
inside a black hole. The second special case is when the gauge field
decouples, i.e. the case
\[
     e \rightarrow  \infty \, .
\]
In this limit, the Maxwell field strength is forced to be pure gauge
and the model reduces to the standard $CP^1$ model except that the
scaling of the kinetic energy term in the Lagrangian corresponds to a
2-sphere of radius $\eta /2$. This is because we have normalized the
action so that the 3-sphere has radius $\eta$ in which case the the
2-sphere of Hopf fibres has radius $\eta  /2$. The metric is now no
longer asymptotically flat but rather asymptotically conical in the
sense described by Barriola and Vilenkin \cite{V&B}, with solid deficit
angle of $4\pi$ times $2\pi G\eta^2$. The case they considered was of
a Higgs field whose vacuum manifold is a 2-sphere of radius $\eta$.
In the corresponding $\sigma$-model limit one can obtain an exact solution
having the metric form (\ref{metricform}), with $Q=0$ and $2\pi G\eta^2$
replaced with $8\pi G\eta^2$. Hence, the solid deficit angle is
four times that in our metric. One can account for this difference
by noting that the energy density falls off as
\[
     {T^0}_0\sim-{{\eta ^2}\over{4r^2}}
\]
in our case and as
\[
     {T^0}_0\sim-{{ \eta ^2}\over{r^2}}
\]
in their case. These expressions differ by a factor of 4, which arises
geometrically because of the different scalings of the kinetic term in
the Lagrangian described above.
Thus, in the limit $e\rightarrow \infty$, our solution represents a
global monopole hidden inside an asymptotically conical black hole.
For finite values of $e$, as long as $\eta \not= 0$, the metric is
also not asymptotically flat but rather asymptotically conical.

It does not seem to be possible to obtain an analytic solution  in the
case of finite $\lambda$. It is, however, possible to obtain an
asymptotic solution near infinity which has the same qualitative form
except that the quantity $Q$ receives a correction of order
$1 \over\lambda $.

It seems \cite{Gol,gary} that global monopoles, inside or outside
black  holes, are  unstable to the  formation of strings, though the
details of this process are not yet clear. In flat space this angular
collapse is prevented by the presence of an $SU(2)$ gauge field in the
case of the 't Hooft-Polyakov  monopole. Moreover, adding the $SU(2)$
gauge field allows the Higgs field to be covariantly constant  near
infinity and hence the monopole to have finite energy. In our case we
only have a $U(1)$ gauge field so that not all of the Higgs fields can
be covariantly constant near infinity and still provide a
non-trivial map from the 2-sphere at infinity to the base of the Hopf
fibration in the $S^3$ target space. Thus our monopole-like solution
still has infinite energy, giving rise to an asymptotically
conical spacetime. It would nevertheless still be interesting to know
whether the presence of the gauge field in our solution can stabilize
it against angular collapse to forming a string.

\section*{Acknowledgements}

We would like to thank Ana Ach{\'u}carro, Mark Hindmarsh, Nick Manton and
Tanmay Vachaspati for helpful discussions.
MEO wishes to thank the SERC for financial support.
FRR was supported by The
Commission of the European Communities through contract No. SC1000488
with The Niels Bohr Institute.

\end{document}